\documentclass[usenatbib]{mn2e}
\usepackage{graphicx}

\newcommand{\msun}{${\rm M_{\sun}}$}

\def\ltsima{$\; \buildrel < \over \sim \;$}
\def\simlt{\lower.5ex\hbox{\ltsima}}
\def\gtsima{$\; \buildrel > \over \sim \;$}
\def\simgt{\lower.5ex\hbox{\gtsima}}
\def\kms{{\rm\,km\,s^{-1}}}

\def\kpc{{\rm\,kpc}}

\def\msun{{\rm\,M_\odot}}

\def\pc{{\rm\,pc}}

\newcommand{\fmmm}[1]{\mbox{$#1$}}
\newcommand{\scnd}{\mbox{\fmmm{''}\hskip-0.3em .}}

\newcommand{\mcnd}{\mbox{\fmmm{'}\hskip-0.3em .}}
\newcommand{\mcnp}{\mbox{\fmmm{'}}}

\def\s{\ifmmode \widetilde \else \~\fi}
\def\={\overline}

\def\spose#1{\hbox to 0pt{#1\hss}}

\def\lta{\mathrel{\spose{\lower 3pt\hbox{$\mathchar"218$}}
     \raise 2.0pt\hbox{$\mathchar"13C$}}}
\def\gta{\mathrel{\spose{\lower 3pt\hbox{$\mathchar"218$}}
     \raise 2.0pt\hbox{$\mathchar"13E$}}}
\def\Dt{\spose{\raise 1.5ex\hbox{\hskip3pt$\mathchar"201$}}}    
\def\dt{\spose{\raise 1.0ex\hbox{\hskip2pt$\mathchar"201$}}}    

\def\dotsfill{\leaders\hbox to 1em{\hss.\hss}\hfill}

\loadboldmathitalic 
\title[A near zero velocity dispersion component in the
CVn dSph]
{A near zero velocity dispersion stellar component in the
Canes Venatici dwarf spheroidal galaxy$^1$}
\author[R. Ibata, S. Chapman, M. Irwin, G. Lewis, N. Martin]
{R. Ibata$^{2}$, S. Chapman$^{3}$, M. Irwin$^{4}$, 
G. Lewis$^{5}$, N. Martin$^{2}$\\
$^{2}$
Observatoire de Strasbourg, 11, rue de l'Universit\'e, F-67000, Strasbourg, 
France\\
$^{3}$
California Institute of Technology, Pasadena, CA 91125, U.S.A\\
$^{4}$
Institute of Astronomy, Madingley Road, Cambridge, CB3 0HA, U.K.\\
$^{5}$
Institute of Astronomy, School of Physics, A29, University of Sydney, NSW
2006, Australia}

\date{\today}
\begin{document} 
\maketitle 
\begin{abstract}
We present a spectroscopic survey of the newly-discovered Canes Venatici
dwarf galaxy using the Keck/DEIMOS spectrograph.  Two stellar populations of
distinct kinematics are found to be present in this galaxy: an extended,
metal-poor component, of half-light radius $7\mcnd8^{+2.4}_{-2.1}$, which
has a velocity dispersion of $13.9^{+3.2}_{-2.5} \kms$, and a more
concentrated (half-light radius $3\mcnd6^{+1.1}_{-0.8}$) metal-rich
component of extremely low velocity dispersion.  At 99\% confidence, the
upper limit to the central velocity dispersion of the metal-rich population
is $1.9\kms$. This is the lowest velocity dispersion ever measured in a
galaxy. We perform a Jeans analysis on the two components, and find that the
dynamics of the structures can only be consistent if we adopt extreme (and
unlikely) values for the scale length and velocity dispersion of the
metal-poor population.  With a larger radial velocity sample and improved
measurements of the density profile of the two populations, we anticipate
that it will be possible to place strong constraints on the central
distribution of the dark matter in this galaxy.
\end{abstract}


\section{Introduction}

\footnotetext[1]{ The data presented herein were obtained at the W.M. Keck
Observatory, which is operated as a scientific partnership among the
California Institute of Technology, the University of California and the
National Aeronautics and Space Administration. The Observatory was made
possible by the generous financial support of the W.M. Keck Foundation.}

Kinematical analyses of the dwarf satellite galaxies show them to be the
astrophysical structures with the highest measured fraction of dark matter,
with some claims of the fraction of total to luminous matter exceeding $M/L
\sim 500$ \citep{kleyna05}.  Until recently, only 11 such dwarf galaxies
were known to orbit the Milky Way, however analysis of the Sloan Digital Sky
Survey (SDSS) has revealed the presence of several lower luminosity dwarf
galaxies that had remained undetected due to their extremely low surface
brightness.  Among the newly-discovered dwarf spheroidal galaxies (dSph)
uncovered by the SDSS is Canis Venatici (CVn), a small structure located at
a distance of approximately $220\kpc$ \citep{zucker06}.  The CVn dSph has an
absolute magnitude of $M_V = -7.9 \pm 0.5$, similar to the Draco dSph
($M_V=-8.8$, \citealt{mateo}), a Milky Way satellite galaxy $71\kpc$ distant
\citep{odenkirchen} that provides us with one of the strongest cases for
containing an extremely large fraction of dark matter. CVn is somewhat more
extended than Draco however, possessing a half-light radius of $550\pc$ (or
$8\mcnd5$, \citealt{zucker06}), compared to $200\pc$ ($9\mcnd7$,
\citealt{wilkinson}) for Draco.  Although Draco is one of the most
convincing examples that dwarf galaxies possess huge mass-to-light ratios,
the tidal effects of the Milky Way on this galaxy are a significant concern
for analyses of its dark matter content, and the dark matter properties that
have been deduced from this galaxy are contingent on the system not being
significantly tidally perturbed.  For this reason the CVn dwarf is an
interesting candidate for a follow-up study to understand the dark
matter. It is similar to Draco, but more distant, and can be expected
therefore to suffer less from the Galactic tides, a reasonable assertion
supported by the larger half mass radius.  Here we present a first
spectroscopic study aimed to reveal the dark matter properties of this
intriguing galaxy.

\section{Observations}

Target stars for observation with DEIMOS were drawn from the SDSS using
simple colour-magnitude criteria designed to select from the upper red giant
branch and extending down to the horizontal branch at $i\sim 22$, as
displayed in Fig.~1. The DEIMOS instrument at the Keck II 10m telescope is a
spectrograph capable of observing up to several hundred targets
simultaneously over a field of $16\arcmin \times 5\arcmin$. On the nights of
May 27 and May 28 2006 we employed this instrument to observe two fields in
the CVn dSph, with exposures of $3\times 20$~min in each field. The
spectrograph was used with the highest resolution 1200~l/mm grating, giving
access to the spectral region 650--900~nm, and with slits of width
$0\scnd7$, resulting in 0.1~nm resolution spectra.  The spectra were
processed using the standard pipeline developed by the DEEP2 collaboration
\citep{davis03}, with wavelength calibration performed by comparison to a
Th-Ar-Ne-Xe arc-lamp exposure taken immediately after the target
observations in order to ensure the best possible velocity accuracy.  The
signal to noise per pixel for the brightest CVn stars was $S/N>50$,
degrading to $S/N \sim 3$ for targets at $i=22$.

\begin{figure}
\begin{center}
\includegraphics[angle=270,width=\hsize]{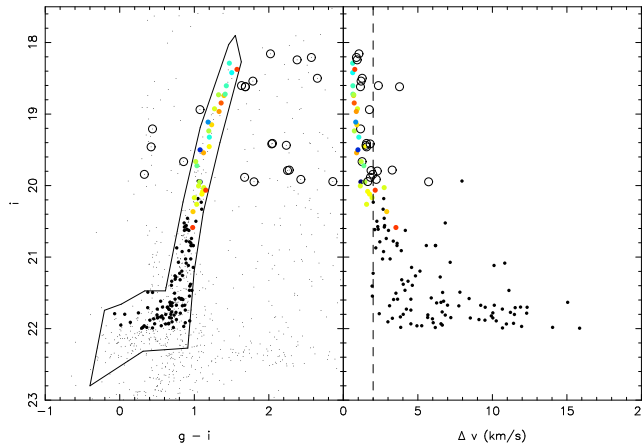}
\end{center}
\caption{The colour-magnitude distribution (CMD) of SDSS stars within
$10\arcmin$ of the galaxy centre is shown on the left panel. Filled circles
mark the locations of the spectroscopically-observed stars that lie in the
selection box, while open circles are low-priority ``filler'' targets.  The
radial velocity uncertainties are displayed in the right hand panel. The
colour of the filled circles indicate the metallicity of the star; the same
coding is used as in Fig.~3.  Only those stars with $S/N > 15$ are
coloured.}
\end{figure}

The velocities of the stars were measured by crosscorrelating the observed
spectra with a synthetic spectrum containing the three prominent absorption
lines of the Ca II ``triplet'', and the corresponding radial velocity
uncertainties were estimated from the rms scatter of measurements to the
three lines fit separately. As shown on the right-hand panel of Fig.~1, the
radial velocity uncertainties for the brightest stars are excellent, being
below $2\kms$ to $i=20$, and rise as expected in fainter targets as the
signal to noise decreases.  We also measure the ${\rm [Fe/H]}$ metallicities
of the stars with the same fitting procedure, via the equivalent width of
the Ca II triplet lines \citep{rutledge}. The metallicities are placed on
the \citet{carretta} scale using the relation ${\rm [Fe/H]} = -2.66 + 0.42
[\Sigma Ca - 0.64 ({\rm V_{HB} - V})]$, with $\Sigma Ca = 0.5 EW_{\lambda
8498} + 1.0 EW_{\lambda 8542} + 0.6 EW_{\lambda 8662}$, ${\rm V_{HB}}$ being
a surface gravity correction relative to the V-band magnitude of the
horizontal branch, and ${\rm V}$ the V-band magnitude of the target CVn
stars. For this, the V-band flux is calculated from SDSS colours using
transformations derived by our group (Ibata et al., in prep.).  We adopt
${\rm V_{HB} = (m-M) - 0.7 = 22.5}$; although this is an approximate
estimate, the derived ${\rm [Fe/H]}$ is not very sensitive to this quantity,
and the relative metallicity measurements are of course unaffected. A
further concern is that $V_{HB}$ is a function of age and metallicity,
though choosing a single value for the mean likely incurs an additional
uncertainty of less than 0.1~dex in ${\rm [Fe/H]}$ \citep{cole04}.  Previous
experience has shown that below $S/N > 15$ (corresponding to $[Fe/H] \sim
0.2$) metallicity measurements become unreliable.  Inspection of the
right-hand panel of Fig.~1 shows that this signal to noise limit corresponds
to a velocity uncertainty of $\Delta v \sim 2\kms$. We therefore adopt $S/N
> 15$ and $\Delta v < 2\kms$ to select our more reliable measurements, which
we label "Sample~A", and which contains 26 stars. In order to test the
statistical limitations of this small sample, we consider a less restricted
selection "Sample~B", with $S/N > 10$ and $\Delta v < 5\kms$, which contains
44 stars.

\begin{figure}
\begin{center}
\includegraphics[angle=270,width=\hsize]{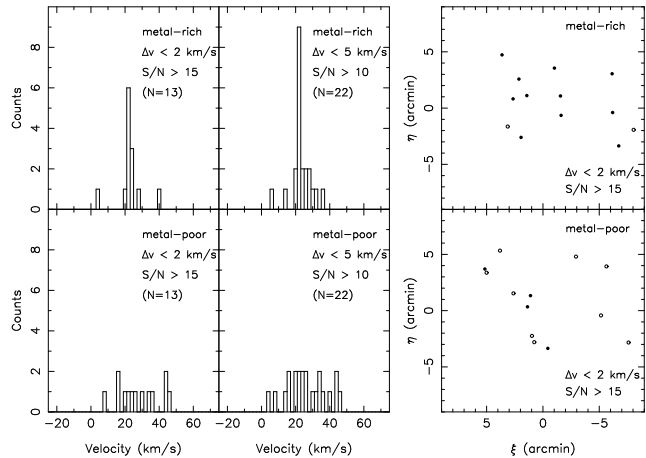}
\end{center}
\caption{The velocity distribution for Samples~A and B are shown on the left
and middle panels, respectively, while the upper and bottom panels display,
respectively, the metal-rich and metal-poor halves of each sample.  The
upper left-hand panel shows the presence of a velocity spike at $22.6\kms$,
with rms dispersion of $1.7\kms$.  The sky positions in Sample~A are shown
on the right-hand panels, with the full circles denoting stars within
$3\sigma$ of the velocity peak.}
\end{figure}

\section{Spectroscopic results}

A first inspection of the data showed a striking dichotomy between the
metal-rich and metal-poor stars.  Dividing Sample~A into a metal-rich half
and a metal-poor half, reveals a narrow peak in the metal-rich sub-sample
that is not present in the metal-poor sub-sample, as shown in the left-hand
panels of Fig.~2. The peak appears to contain 11 stars, and is centred at
$22.6\kms$ with an rms dispersion of $1.7\kms$ (i.e. before removing
instrumental effects). A further 2 stars are present in this selection,
although displaced by more than $10\sigma$ from this narrow peak. As we
relax the maximum velocity uncertainty limit to $\Delta v < 5 \kms$
(Sample~B), the peak is reinforced, as we show in the middle panels of
Fig.~1, though it becomes smeared, consistent with the increased
instrumental uncertainty.  The F-test indicates that the probability that
these two samples have the same variance is very low, $1.8\times10^{-3}$.
This first cursory inspection of our CVn dSph sample therefore indicates
that this galaxy possesses a metal-rich component with very small velocity
dispersion of $1.7\kms$, together with a kinematically hotter component of
lower metallicity.  The stars within the two components are broadly
distributed over the observed area, as demonstrated in the right-hand panels
of Fig.~1.

\begin{figure}
\begin{center}
\includegraphics[angle=270,width=\hsize]{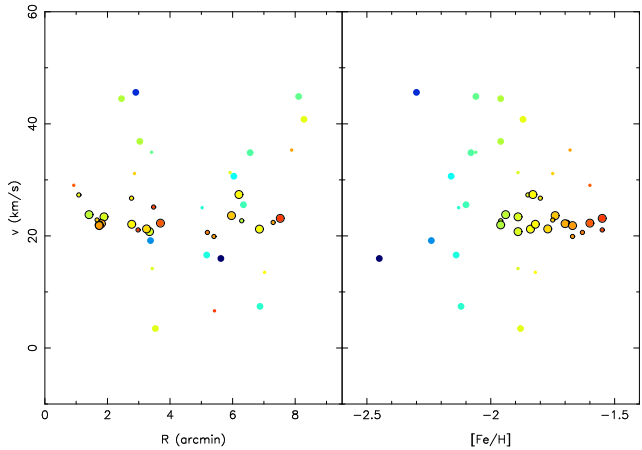}
\end{center}
\caption{The velocity-radius and velocity-metallicity relations for Sample~A
(large dots), and Sample~B (small dots).  The clear grouping of metal-rich
stars with ${\rm [Fe/H] > -2.0}$ and with $17.5 < v < 27.7\kms$ are circled
in this panel. The radial location of these stars is indicated in the
left-hand panel (with the same colour code for metallicity).  The CVn dSph
evidently contains stars with a wide range of metallicities, and includes
stars at the lowest end of the abundance distribution of this class of
galaxies.}
\end{figure}

Given this exceedingly low central velocity dispersion it is relevant to ask
whether these stars are remnants of a disrupted star cluster. The right-hand
panel of Fig.~3 shows the metallicity-velocity relation of Samples A and B.
It is clear from this diagram that the stars that partake in the narrow
velocity peak shown previously in Fig.~2 cover a broad range of metallicity
from ${\rm [Fe/H]=-2.0}$ to ${-1.5}$. This shows conclusively that CVn
became chemically enriched progressively over time, a clear sign of a galaxy
that possesses (or once possessed) sufficient mass to retain and enrich
gas. In contrast, star clusters have simple coeval populations of single
metallicity, with the notable exception of $\omega$~Cen, which is thought to
have been the core of a now disrupted dwarf galaxy (see e.g.,
\citealt{bekki}).

Fig.~3 shows that it is natural to divide CVn into a dynamically-cold,
metal-rich component (circled points), and a dynamically hot, metal-poor
component (the remaining non-circled points).  Such behaviour has previously
been discovered in the Sculptor dSph \citep{tolstoy}.  The boundary between
the subsets was chosen at ${\rm [Fe/H] > -2.0}$ and $17.5 < v < 27.7\kms$
(i.e. $3\sigma$ from the peak in Fig.2); in the discussion below, we refer
to the stars within this selection region as the "metal-rich component" and
those outside as the "metal-poor component".  We use a maximum-likelihood
technique \citep{fisher1922} to correct for the individual radial velocity
uncertainty estimates and thereby measure the intrinsic velocity dispersions
of these components. The 13 Sample~A stars belonging to the metal-rich
selection display a velocity dispersion of $0.5\kms$ with a 99\% upper limit
of $1.9\kms$ (the corresponding likelihood contours are displayed in red on
the right-hand panel in Fig.~4).  In contrast, the 13 Sample~A stars in the
metal-poor selection are found to have $\sigma_v = 13.9^{+3.2}_{-2.5} \kms$
(likelihood contours displayed dark blue in Fig.~4).  With the looser
selection criteria of Sample~B, we measure $\sigma_v = 0.5\kms$, with an
upper limit of $1.8\kms$ (99\% confidence) for the metal-rich selection, and
$\sigma_v = 12.5^{+2.1}_{-1.8} \kms$, with a lower limit of $8.7\kms$ (99\%
confidence) for the metal-poor selection. Essentially identical results are
therefore obtained from the different quality cuts of Samples A and B.

\begin{figure}
\begin{center}
\includegraphics[angle=270,width=\hsize]{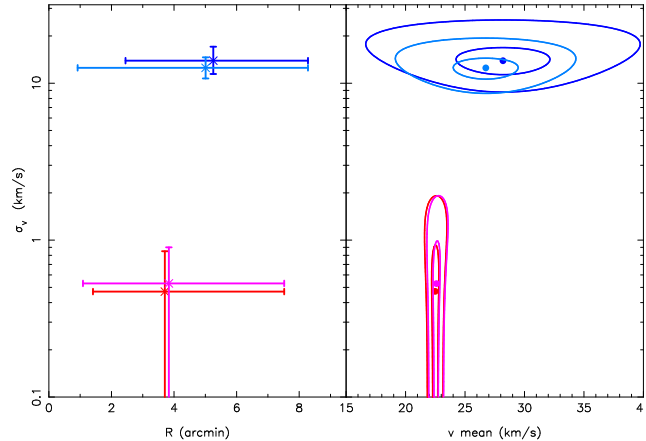}
\end{center}
\caption{The likelihood contours (at 68.3\% and 99\%) of a Gaussian fit to
the stellar sample displayed in Fig.~3 are shown in the right-hand panel, as
a function of velocity dispersion and mean velocity. Dark red and dark blue
correspond, respectively, to the metal-rich and metal-poor selections from
Sample~A, while pink and light blue show, respectively the results from the
metal-rich and metal-poor selections from Sample~B.  The mean radial
location of the samples are shown in the left-hand panel; here the vertical
bars show the $1\sigma$ velocity dispersion uncertainty, while the
horizontal bars show the full radial range in each sub-sample.}
\end{figure}

 \section{Structure of CVn revisited}
 
We have attempted a reanalysis of the spatial structure of CVn using SDSS
photometry.  Given the strong chemical differences in the two kinematic
components, we hoped to be able to disentangle the two populations using the
colour of the red-giant branch (RGB) stars.  Unfortunately, no significant
differences were discovered dividing the RGB sample into blue and red
sub-samples. The reason for this failure is evident from an inspection of
the left-hand panel of Fig.~1, which demonstrates that there is a very small
difference in RGB colour between stars of very different
(spectroscopically-measured) metallicity, indicating that the metal-rich
stars are substantially younger than the metal-poor stars. An alternative
diagnostic that has proved to be useful in previous analyses of dSph is the
comparison between the blue and red horizontal branches (HB)
\citep{tolstoy}. To this end we divided the CMD selection box of Fig.~1 into
a blue HB (stars with $g-i < 0.2$), a red HB sample (all stars with $0.2 <
g-i < 0.6$) and an RGB sample (all remaining stars in the CMD selection
polygon). The corresponding contours of star density are displayed in
Fig.~5. Though the blue and red HBs are very similar it is surprising to
find their centres displaced from that of the RGB sample.  Inspection of
deeper Subaru/Suprimecam data, for which unfortunately only i-band images
were available, reveal that approximately half of the sources in the SDSS HB
peak are in fact due to faint galaxies which are detected as unresolved
point-sources in the SDSS. However, the correction for this contamination
does not make the HB peak disappear entirely. Unfortunately, the Subaru data
were taken at very high airmass and do not sample the galaxy uniformly. To
prove or dismiss the reality of the offset HBs (itself a very puzzling
phenomenon!) will require better quality data.

\begin{figure}
\begin{center}
\includegraphics[angle=270,width=\hsize]{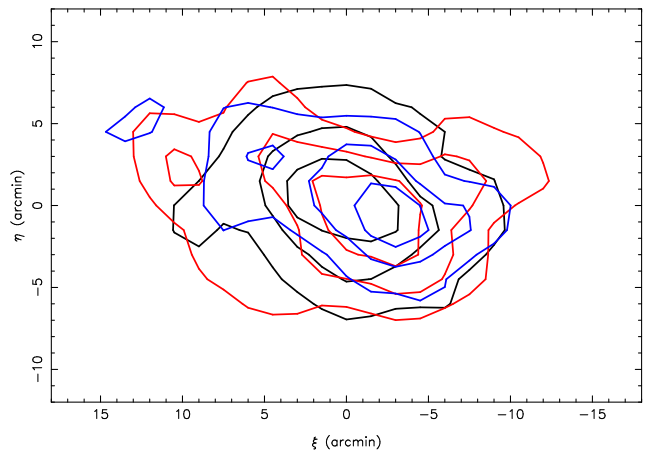}
\end{center}
\caption{The SDSS stellar density contours of the RGB sample (black), the
blue HB sample (blue) and the red HB sample (red).}
\end{figure}

\begin{figure}
\begin{center}
\includegraphics[angle=270,width=\hsize]{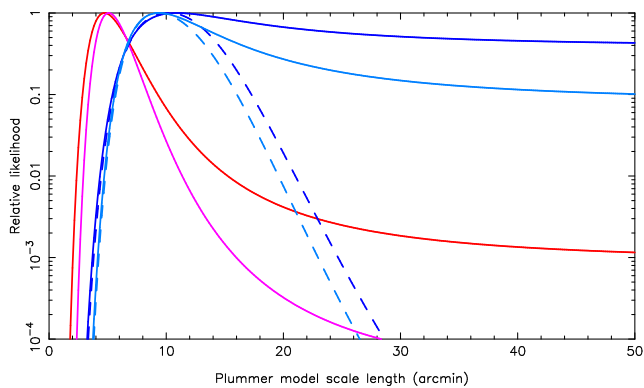}
\end{center}
\caption{The relative likelihood of Plummer model fits to the metal-rich
(red and pink) and metal-poor (dark blue and light blue) samples, as a
function of the Plummer model scale parameter $a$. The same colour coding is
used as in Fig.~4. Full lines use only the spectroscopically observed stars,
while dashed lines complement these with SDSS photometry.}
\end{figure}

Despite the fact that the SDSS imaging data do not allow us to separate the
radial profiles of the two kinematic populations, we can attempt to
constrain these profiles using the DEIMOS dataset. All of the bright stars
with $S/N>10$ within the selection box of Fig.~1 were chosen by the DEIMOS
slitlet configuration program with identical priorities. Assuming that the
target selection program does not bias object allocation with radius, our
sample can be considered to trace the two populations within the spatial
``window function'' of the two DEIMOS fields.  We fit a Plummer model to the
samples previously displayed in Fig.~4, using a maximum-likelihood method.
The relative likelihood of the value of the Plummer scale length $a$ is
displayed in Fig.~6 (note that the conversion to half-mass radius is $r_h =
a \sqrt{2^{2/3}-1}$).  For Sample~A, the most likely value for the
metal-rich component (shown in red) is $a=4\mcnd7$, while the metal-poor
component (shown in a continuous dark blue line) is much more extended, with
the most likely value of $a$ being $10\mcnd9$, though much larger values of
$a$ are also acceptable.  Of course the reason that very large values of $a$
are acceptable is simply that the kinematic sample is confined to
$R<9\arcmin$. To improve the constraint on $a$ we supplement the metal-poor
star sample with RGB stars detected in the SDSS, at radii between $9\arcmin$
(i.e., twice the scale length of the metal-rich component) and $24\arcmin$
(where the profile disappears into the background). The corresponding
likelihood curve is displayed in Fig.~6 with a dashed dark blue line, and
possesses a maximum at $a=10\mcnd2$, similar to the result from the
kinematic sample alone, but falls off rapidly for large $a$.  Very similar
results are derived for Sample~B (see Fig.~6).  We therefore find that the
metal-rich population is more centrally concentrated than the metal-poor
population.

\begin{figure}
\begin{center}
\includegraphics[angle=270,width=\hsize]{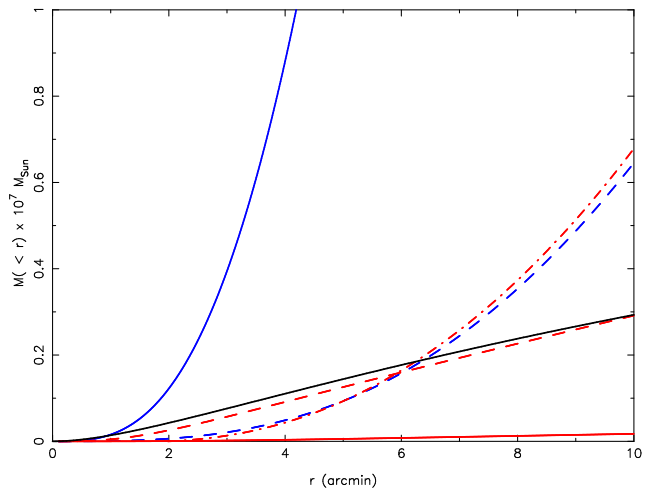}
\end{center}
\caption{The enclosed mass as a function of radius.  Results for the
metal-poor sub-component are shown in blue.  The full line shows the
preferred model, which has an isothermal dispersion of $\sigma_v = 13.9\kms$
and Plummer scale length of $a=10\mcnd2$, while the dashed line is an
attempt to minimise the enclosed mass and has $\sigma_v = 6.4\kms$ and
$a=21\arcmin$.  The red lines are the mass profiles derived from the
metal-rich population.  The full line shows again the consequence of
adopting the most likely parameters: an isothermal velocity dispersion of
$\sigma_v = 0.5\kms$ and a Plummer scale parameter of $a=4\mcnd7$, while the
dashed line shows an extreme model having $\sigma_v = 1.9\kms$ and
$a=2\mcnd4$.  The dot-dashed line is an attempt to reconcile the mass
profile of the metal-rich sub-system with the metal-poor sub-system: it
possesses the preferred Plummer scale length of $a=4\mcnd7$, but has
linearly increasing velocity dispersion profile with a central value of
$\sigma_v = 1.5\kms$, increasing to $\sigma_v = 3.2\kms$ at $6\mcnd6$.  For
comparison, we superimpose the mass profile of a \citet{navarro} model of
Virial mass $10^7\msun$ and concentration $c=20$ (black line).}
\end{figure}

 \section{Discussion}

The upper limit to the velocity dispersion of the metal-rich component in
CVn is the lowest ever measured in a galaxy. As such, it presents a
fascinating problem for galaxy formation. The simplest explanation may be
that CVn was a normal dark-matter dominated dSph that has been severely
tidally disrupted over time.  The removal of the extended dark matter halo
of the galaxy by tidal forces could lower the central velocity dispersion of
the stellar component by lowering the mass of the system.  In this
interpretation, the higher dispersion in the external regions would attest
to on-going tidal disruption of the outermost stars. However, the regular
contours found by \citet{zucker06} are suggestive of only mild tidal forces,
which argues against this scenario.

Assuming that the system is in dynamical equilibrium, we may apply the Jeans
equations to deduce the mass profile of the dwarf galaxy, including the dark
matter.  With the further assumptions that the galaxy is spherically
symmetric, that the two stellar components follow the (projected) Plummer
density distributions fitted previously, and that there is no appreciable
velocity anisotropy, the Jeans equations can be simplified to Eqns. 4-56 and
4-58 of \citet{BT}.  The solid blue and red lines in Fig.~7 show the
computed mass profiles for the metal-poor and metal-rich components,
respectively, using the most likely model parameters.  For this we have
taken the parameters $a=10\mcnd2$ and $\sigma_v=13.9\kms$ (metal-poor), and
$a=4\mcnd7$ and $\sigma_v=0.5\kms$ (metal-rich), with the further assumption
that the velocity dispersion are constant at all radii. The two derived mass
profiles could hardly be more different, and are clearly completely
inconsistent with each other.  To obtain a degree of consistency between the
components (dashed lines), we are obliged to force the metal-poor component
to $a=21\arcmin$, $\sigma_v=6.4\kms$ (the 99\% upper and lower limits for
these parameters, respectively), and take the opposite limits for the
metal-rich component: $a=2\mcnd4$, $\sigma_v=1.9\kms$.  The combination of
these extreme parameter values is highly unlikely of course, and illustrates
the extent to which the two components are apparently inconsistent.
However, without a more reliable description of the components it is
difficult to be categorical about the inconsistency. In particular, we have
little information on the radial profile of the velocity dispersion. The
effect of this uncertainty is illustrated by the dot-dashed line, where we
have taken the most likely Plummer scale parameter for the metal-rich
structure, $a=4\mcnd7$, together with a linearly-increasing velocity
dispersion model that has a central value of $1.5\kms$, increasing to
$3\kms$ at $6\mcnp$.  With these parameters the mass profile required by the
metal-rich component agrees very well with the extreme configuration of the
metal-poor structure (blue dashed line).


\section{Conclusions}

We have uncovered two stellar components in the Canes Venatici dwarf
spheroidal galaxy with distinct structure and kinematics.  Below a
metallicity of ${\rm [Fe/H] = -2}$ this galaxy has a velocity dispersion of
$\sim 10\kms$, typical of other dSph galaxies. At metallicities richer than
this limit, an intriguing additional component is detected, possessing a
velocity dispersion that is barely resolved with our measurement accuracy,
and has an upper limit of $1.9\kms$. The nature of the metal-rich component
is unclear, though the wide metallicity spread it displays suggests that it
was formed in situ from gas reprocessed from the previous generation of
metal-poor stars. The possibility that the metal-rich population is younger
is supported by the overlap of the red-giant branches of both populations in
the CMD.

Thus the CVn dwarf presents us with a very interesting puzzle: why does this
younger population possess such a small velocity dispersion?  Naively, one
would expect this to attest to the presence of a cored potential in the
central regions of the galaxy: where the potential gradients are flat one
can hide much dark mass without structures that are significantly smaller in
scale than the core radius being substantially affected.  However, the Jeans
analysis we have performed does not support this scenario, due to the
relatively small scale length of the metal-poor population.  Instead, we
find that the model parameters must be forced to highly unlikely extremes to
achieve a reasonable consistency between the components. We feel, however,
that the paucity of well-measured stars in the kinematic sample, especially
at large radius, introduces a large modelling uncertainty, and we therefore
avoid stating categorically at the present time that the two populations are
inconsistent.

If future studies confirm the values for the velocity dispersion and
component scale lengths presented here, it will be a clear indication that
one or more of the assumptions made in the Jeans analysis is invalid. The
system may not be in dynamical equilibrium: despite the previous discussion,
it is possible that the metal-rich population is a recent accretion that has
not had time to relax. Given typical dynamical times, this seems very
unlikely, though we note that in other dSph fossil substructure has been
detected, and its longevity can be interpreted as consistent with the
presence of a cored dark matter distribution
\citep{kleyna03}. Alternatively, the adopted assumption of spherical
symmetry may be poor. The metal-rich component may also be a disk-like
component seen face-on, though this is hard to envisage if it was formed
from the existing apparently non-rotating metal-poor population, given the
similarity in their scale lengths.  It may also turn out to be that the
velocity distribution is not isotropic.

Although the present uncertainties in the model parameters do not allow us
to provide strong constraints on the dark matter distribution, we anticipate
that if the predictions of the two components can be made consistent with
improved data, this will yield a reliable dark matter profile, from which we
may confidently determine whether the dark matter conforms to the
predictions of CDM theory or not.

%

\newcommand{\mnras}{MNRAS}
\newcommand{\nat}{Nature}
\newcommand{\araa}{ARAA}
\newcommand{\aj}{AJ}
\newcommand{\apj}{ApJ}
\newcommand{\apjl}{ApJ}
\newcommand{\apjs}{ApJSupp}
\newcommand{\aap}{A\&A}
\newcommand{\aaps}{A\&ASupp}
\newcommand{\pasp}{PASP}

\end{document}